\documentclass[11pt]{article}
\usepackage[dvips]{graphicx}
\usepackage{graphics}
\usepackage{graphicx}
\usepackage{amssymb}
\usepackage{amsmath}
\usepackage{amssymb}
\usepackage{overcite}
\usepackage{epsf}
\input{epsf}
\numberwithin{equation}{section}
\begin{document}
\begin{titlepage}
\title{\bf Bogoliubov's Quasiaverages,  Broken Symmetry and Quantum Statistical Physics}  
%
\author{
A. L. Kuzemsky \thanks {E-mail:kuzemsky@theor.jinr.ru;
http://theor.jinr.ru/\symbol{126}kuzemsky}
\\
{\it Bogoliubov Laboratory of Theoretical Physics,} \\
{\it  Joint Institute for Nuclear Research,}\\
{\it 141980 Dubna, Moscow Region, Russia.}}
\date{}
\maketitle
\begin{abstract}
The development and applications of the method  of quasiaverages developed by N. N. Bogoliubov to 
quantum statistical physics and
to quantum solid state theory and, in particular, to quantum theory of magnetism,  were  analyzed.
The problem of finding  the ferromagnetic, antiferromagnetic and superconducting
\emph{symmetry broken}  solutions of the correlated lattice fermion
models was discussed  within the irreducible Green functions method. A
unified scheme for the construction of generalized mean fields (elastic scattering corrections) 
and self-energy (inelastic scattering) in terms of the Dyson equation was  generalized in
order to include  the \emph{source fields}. The interrelation of the Bogoliubov's idea 
of quasiaverages and  the concepts of symmetry breaking and quantum protectorate was discussed briefly  in the context 
of quantum  statistical physics. The idea of quantum protectorate reveals the essential
difference in the behavior of the complex many-body systems at the low-energy
and high-energy scales. It was shown that the role of symmetry (and the breaking of symmetries) in combination 
with the degeneracy of the system was reanalyzed and essentially clarified within the framework of the method  
of quasiaverages. The complementary notion of quantum protectorate might provide
distinctive signatures and good criteria for a hierarchy of energy scales and  the appropriate emergent behavior.

%
%
\end{abstract}
\end{titlepage}
%
%

%
\section{  Introduction}
%
\noindent
It is well known that symmetry principles play a crucial role in physics~\cite{pwa84,rose}. The theory of symmetry is
a basic tool for understanding and formulating the fundamental notions of physics~\cite{baro}.
According to F. Wilczek~\cite{wil05},  "the primary goal of fundamental physics is to discover profound concepts that 
illuminate our understanding of nature".
It is known  that    symmetry is a driving force in the shaping of physical theory;
moreover, the primary lesson of physics of last century is that the secret of nature is symmetry.
Every symmetry leads to a conservation law;  the well known examples are the conservation of
energy, momentum and electrical charge. A variety of other conservation laws can be deduced from symmetry
or invariance properties of the corresponding Lagrangian or  Hamiltonian of the system.  According to Noether
theorem, every continuous symmetry transformation under which the Lagrangian of a given system remains invariant
implies the existence of a conserved function~\cite{rose}.  
Many fundamental laws of physics in addition to their detailed features possess various symmetry properties. These
symmetry properties lead to  certain constraints and regularities on the possible properties of matter. 
Thus the  principles of symmetries  belong to the \textbf{underlying principles} of physics. \\ 
It is known that when the Hamiltonian of a system is invariant under a symmetry operation, but the
ground state is not, the symmetry of the system can be spontaneously broken~\cite{fstr05}. 
Symmetry breaking is termed \emph{spontaneous} when there is no explicit term in a Lagrangian which
manifestly breaks the symmetry. 
Symmetries and  breaking of symmetries play an important role in  statistical physics, quantum field theory, physics
of elementary particles, etc.~\cite{namb07,namb09}\\
In physics, spontaneous symmetry breaking occurs when a system that is symmetric with respect to some symmetry 
group goes into a vacuum state that is not symmetric. When that happens, the system no longer appears to behave in 
a symmetric manner. It is a phenomenon that naturally occurs in many situations.
The mechanism of spontaneous symmetry breaking is usually understood as the mechanism
responsible for the occurrence of asymmetric states in quantum systems in the thermodynamic limit and is used
in various field of quantum physics~\cite{grib}.  
The intriguing mechanism of spontaneous symmetry breaking is a unifying concept that lie at the basis of most 
of the recent developments in theoretical physics, from statistical mechanics to many-body theory and to elementary 
particles theory~\cite{namb07,namb09}.\\ 
It should be stressed that symmetry implies degeneracy. The greater the symmetry, the greater the degeneracy.
The study of the degeneracy of the energy levels plays a very important role in quantum physics. 
It is of importance to emphasize that when spontaneous symmetry breaking takes place, the ground state of the system
is degenerate. 
Substantial progress in the understanding of the  broken symmetry concept was connected with
Bogoliubov's fundamental   ideas on quasiaverages~\cite{nnb60,nnb61,nnb61w,nnb71,bb09,nnbj72,nnbj,bs75,nnbj3,petr95}. 
Studies of degenerate systems led Bogoliubov in 1960-61 to the formulation of \textbf{the method of quasiaverages}.
This method has proved to be a universal tool for systems whose ground states become unstable under small
perturbations. 
Thus the role of symmetry (and the breaking of symmetries) in combination 
with the degeneracy of the system was reanalyzed and essentially clarified by N. N. Bogoliubov in 1960-1961. He 
invented and formulated
a powerful innovative idea of \emph{quasiaverages} in statistical mechanics~\cite{nnb60,nnb61,bb09,nnbj,bs75,petr95}.
The very elegant work of N. N. Bogoliubov~\cite{nnb61} has been
of great importance for a deeper understanding of phase  transitions, superfluidity  and 
superconductivity, quantum theory of magnetism~\cite{tyab} and other fields of equilibrium and nonequilibrium
statistical mechanics~\cite{nnb61,nnb61w,nnb71,bb09,nnbj,bs75,petr95,zub71,kuz07,kuz09,kuz10}. 
The concept of quasiaverages is indirectly related to the theory of phase transition. The instability of thermodynamic 
averages with respect to perturbations of the Hamiltonian by a breaking of the invariance with respect to a certain 
group of transformations means that in the system transition to an extremal state occurs.
The mathematical apparatus of the method of quasiaverages includes the Bogoliubov theorem~\cite{nnb61,bb09,petr95,sado73} on 
singularities of type $1/q^{2}$  and the Bogoliubov 
inequality for Green and correlation functions as a direct consequence of the method. It includes algorithms 
for establishing non-trivial estimates for equilibrium quasiaverages, enabling one to study the problem of ordering in statistical systems 
and to elucidate the structure of the energy spectrum of the underlying excited states. 
Thus the  Bogoliubov's idea  of \emph{quasiaverages} is an essential
conceptual advance of modern physics. \\
Many-particle systems where the interaction 
is strong have often complicated behavior, and require nonperturbative approaches to treat their properties.  
Such  situations  are often arise in condensed matter systems.
Electrical, magnetic and mechanical properties of materials are \emph{emergent collective behaviors}  of the underlying quantum mechanics of 
their electrons and constituent atoms. A principal aim of solid state physics and materials science is to elucidate this emergence. 
A full achievement of this goal would imply the ability to engineer a material that is optimum for any particular 
application. The current understanding of electrons in solids uses simplified but workable picture
 known as the Fermi liquid theory. This theory explains why 
electrons in solids can often be described in a simplified manner which appears to ignore the large repulsive forces 
that electrons are known 
to exert on one another. There is a growing appreciation that this theory probably fails for entire classes of possibly 
useful materials and 
there is the suspicion that the failure has to do with unresolved competition between different possible emergent 
behaviors.\\ 
It is appropriate to note here that the emergent properties of matter were analyzed and discussed
by R. Laughlin and D. Pines~\cite{pnas,pines} from a general point of view (see also Ref.~\cite{wen05}). 
They introduced a unifying idea of \emph{quantum protectorate}.
This concept  belongs also to the underlying principles  of physics.   
The idea of quantum protectorate reveals the essential
difference in the behavior of the complex many-body systems at the low-energy
and high-energy scales.  The existence of two
scales, low-energy and high-energy, in the description
of physical phenomena is used in physics, explicitly or implicitly.\\ 
From the other hand,
it was
recognized for many years that  the strong
correlation in solids exist between the motions of various
particles (electrons and ions, i.e.  the fermion and boson
degrees of freedom) which arise from the Coulomb forces. The most
interesting objects are metals and their compounds. They are
invariant under the translation group of a crystal lattice and
have lattice vibrations as well as electron degrees of freedom.
There are many evidences for the importance of  many-body effects
in these systems. Within the  semi-phenomenological theory
it was suggested that the low-lying excited states of an
interacting Fermi gas can be described in terms of a set of {\it
"independent quasiparticles"}. However, this was a
phenomenological approach and did not reveal the nature of
relevant interactions.
An alternative way of viewing quasiparticles,  more general and
consistent, is through the Green function scheme of many-body
theory~\cite{tyab,akuz09,kuz09,bt59}.
It becomes clear that only a thorough experimental and theoretical investigation of quasiparticle many-body dynamics
of the many particle systems can provide the answer on the relevant microscopic picture~\cite{kuz09}.
In our works, we  discussed the microscopic view of a dynamic
behavior of various interacting many-body systems on a lattice~\cite{kuz09,kuz78,kuz89,km90,kuz94,ckw96,kuz96,kuzsf99,kuz99,rnc02,kuz04,kuz05}. 
A comprehensive description of transition and rare-earth metals and alloys and other materials 
(as well as efficient predictions of properties
of new materials) is possible only in those cases, when
there is an adequate quantum-statistical theory based
on the information about the electron and crystalline
structures. The main theoretical problem of this direction
of research, which is the essence of the
quantum theory of magnetism, is investigations and improvements
of quantum-statistical models describing the
behavior of the complex compounds and materials in order
to take into account the main features of their electronic
structure, namely, their dual "band-atomic" nature~\cite{kuz09}. 
The construction of a consistent theory explaining
the electronic structure of these substances encounters
serious difficulties when trying to describe the collectivization-localization duality
in the behavior of electrons.
This problem appears to be extremely important,
since its solution gives us a key to understanding magnetic,
electronic, and other properties of this diverse
group of substances. The author of the present paper
investigated the suitability of the basic models with
strong electron correlations and with a complex spectrum
for an adequate and correct description of the dual
character of electron states~\cite{kuz09}. A universal mathematical
formalism was developed for this investigation~\cite{kuz09,rnc02}. It
takes into account the main features of the electronic
structure and allows one to describe the true quasiparticle
spectrum, as well as the appearance of the magnetically
ordered, superconducting, and dielectric (or
semiconducting) states.
With a few exceptions, diverse physical phenomena
observed in compounds and alloys of transition and
rare-earth metals, cannot be explained in
the framework of the mean-field approximation, which
overestimates the role of inter-electron correlations in
computations of their static and dynamic characteristics.
The circle of questions without a precise and definitive
answer, so far, includes such extremely important
(not only from a theoretical, but also from a practical
point of view) problems as the adequate description of
quasiparticle dynamics for quantum-statistical models
in a wide range of their parameter values. The source of
difficulties here lies not only in the complexity of calculations
of certain dynamic properties (such as, the
density of states, electrical conductivity, susceptibility,
electron-phonon spectral function, the inelastic scattering
cross section for slow neutrons), but also in the
absence of a well-developed method for a consistent
quantum-statistical analysis of a many-particle interaction
in such systems. A self-consistent field approach
was used in the papers~\cite{kuz09,kuz78,kuz89,km90,kuz94,ckw96,kuz96,kuzsf99,kuz99,rnc02,kuz04,kuz05} 
for description of various dynamic characteristics of strongly correlated
electronic systems. It allows one to consistently and
quite compactly compute quasiparticle spectra for
many-particle systems with strong interaction taking
into account damping effects. The correlation effects
and quasiparticle damping are the determining factors
in analysis of the normal properties of high-temperature
superconductors, heavy fermion compounds, etc.
 We also formulated a
general scheme for a theoretical description of electronic
properties of many-particle systems taking into
account strong inter-electron correlations~\cite{kuz09,rnc02}.
The scheme is a synthesis of the method of two-time
temperature Green's functions~\cite{tyab,bt59} and the diagram
technique. An important feature of this approach is a
clear-cut separation of the elastic and inelastic scattering
processes in many-particle systems (which is a
highly nontrivial task for strongly correlated systems).
As a result, one can construct a correct basic approximation
in terms of generalized mean fields (the elastic
scattering corrections), which allows one to describe
magnetically ordered or superconducting states of the
system. The residual correlation effects, which are the
source of quasiparticle damping, are described in
terms of the Dyson equation with a formally exact representation
for the mass operator.\\  
In the present paper we will discuss some
applications of the symmetry principles to quantum and statistical physics and quantum solid state theory
in the light of our results on quasiparticle many-body dynamics.
%
%
%
%
\section{Bogoliubov's Quasiaverages in Statistical Mechanics}
%
%
In the work of N. N. Bogoliubov
"Quasiaverages in Problems of Statistical Mechanics" the innovative notion of \emph{quasiaverege}~\cite{nnb61}
was introduced and applied to various problem of statistical physics. In particular,
quasiaverages of Green's functions constructed from ordinary averages, degeneration 
of statistical equilibrium states, principle of weakened correlations, and particle pair states were considered. 
In this framework the $1/q^{2}$-type properties in the theory of the superfluidity of Bose and Fermi systems, the properties of basic 
Green functions for a Bose system in the presence of condensate, and a model with separated condensate 
were analyzed.\\
The method of quasiaverages is a constructive workable scheme for studying systems with spontaneous symmetry breakdown. 
A quasiaverage is a thermodynamic (in statistical mechanics) or vacuum (in quantum field theory) average of dynamical 
quantities in a specially modified averaging procedure, enabling one to take into account the effects of the influence of 
state degeneracy of the system.
The method gives the so-called macro-objectivation of the degeneracy in the domain of quantum statistical mechanics and in quantum physics.
In statistical mechanics, under spontaneous symmetry breakdown one can, by using the method of quasiaverages, describe 
macroscopic observable within the framework of the microscopic approach.\\
In considering problems of findings the eigenfunctions in quantum mechanics it is well known that the theory of 
perturbations should be modified substantially for the degenerate systems. In the problems of statistical mechanics
we have always the degenerate case due to existence of the additive conservation laws.
The traditional approach to quantum statistical mechanics~\cite{petr95} is based on the unique canonical
quantization of classical Hamiltonians for systems with finitely many degrees of freedom together with the
ensemble averaging in terms of traces involving a statistical operator $\rho$.
For an operator $\mathcal{A}$ corresponding to some physical quantity $A$ the average value of $A$ will be given as
\begin{equation}\label{q1}
\langle A \rangle_{H}  =   \textrm{Tr} \rho A ; \quad  \rho  = \exp ^{- \beta H} / \textrm{Tr} \exp ^{- \beta H},         
\end{equation}
where $H$ is the Hamiltonian of the system, $\beta = 1/kT$ is the reciprocal of the temperature.\\
The core of the problem lies in establishing the existence of a thermodynamic limit (such as $N/V = $ const, 
$V \rightarrow \infty$, $N$ = number of degrees of freedom, $V$ = volume) and its evaluation for the quantities 
of interest. 
Thus in the statistical mechanics the average $\langle A \rangle$ of any dynamical quantity $A$
is defined in a single-valued way. 
In the situations with degeneracy  the specific problems appear. In quantum mechanics, if two linearly independent
state vectors (wavefunctions in the Schroedinger picture) have the same energy, there is a degeneracy.
In this case more than one independent state of the system corresponds to a single energy level.  
If the statistical equilibrium state of the system 
possesses lower symmetry than the Hamiltonian of the system (i.e. the situation with the  symmetry breakdown), 
then it is necessary to supplement the averaging procedure  (\ref{q1}) by a rule forbidding  irrelevant   averaging over the values of macroscopic quantities considered 
for which a change is not accompanied by a change in energy. 
This is achieved by introducing quasiaverages, that is, averages over the Hamiltonian $H_{\nu \vec{e}}$ 
supplemented by infinitesimally-small 
terms that violate the additive conservations laws 
$H_{\nu \vec{e}} = H + \nu (\vec{e}\cdot \vec{M})$, ($\nu \rightarrow 0$). Thermodynamic 
averaging may turn out to be unstable with respect to such a change of the original 
Hamiltonian, which is another indication of degeneracy of the equilibrium state. 
According to Bogoliubov~\cite{nnb61}, the quasiaverage of a dynamical quantity  $A$ for the system with the
Hamiltonian $H_{\nu \vec{e}}$  
is defined as the limit
\begin{equation}
\label{q9}
\curlyeqprec A \curlyeqsucc = \lim_{\nu \rightarrow 0} \langle A \rangle_{\nu \vec{e}},
\end{equation}
where $\langle A \rangle_{\nu \vec{e}}$  denotes the ordinary average taken over the Hamiltonian $H_{\nu \vec{e}}$, containing the small symmetry-breaking terms introduced by the inclusion 
parameter $\nu$, which vanish as $\nu \rightarrow 0$ after passage to the thermodynamic limit $V \rightarrow \infty$. 
It is important to note that in this equation limits cannot be interchanged.
Thus the existence of degeneracy is reflected directly in the quasiaverages by their dependence upon the arbitrary
unit vector $\vec{e}$.
It is also clear that
\begin{equation}
\label{q10}
\langle A \rangle = \int \curlyeqprec A \curlyeqsucc d \vec{e}.
\end{equation}
According to definition (\ref{q10}), the ordinary thermodynamic average is obtained by extra averaging of the 
quasiaverage over the symmetry-breaking group. Thus to describe the case of a degenerate state of statistical
equilibrium quasiaverages are more convenient, more physical, than ordinary averages~\cite{petr95}. The latter are the same 
quasiaverages only averaged over all the directions $\vec{e}$.\\
It is necessary to stress,
that the starting point for Bogoliubov's work~\cite{nnb61} was
an investigation of additive conservation laws and
selection rules, continuing and developing the  approach by P. Curie  for derivation of
selection rules for physical effects (see also Ref.~\cite{namb07}). Bogoliubov demonstrated
that in the cases when the state of statistical
equilibrium is degenerate, as in the case of the Heisenberg ferromagnet,
one can remove the degeneracy of equilibrium
states with respect to the group of spin rotations by
including in the Hamiltonian $H$ an additional noninvariant
term $\nu M_{z} V$ with an infinitely small $\nu$.  For the Heisenberg ferromagnet the ordinary averages must be invariant
with regard to the spin rotation group. The corresponding quasiaverages possess only the property of covariance.
Thus the quasiaverages do not follow the same selection rules as those which govern ordinary averages, due to their 
invariance with regard to the spin rotation group. It is clear that that the unit vector $\vec{e}$, i.e., the
direction of the magnetization $\vec{M}$ vector, characterizes the degeneracy of the considered state of statistical
equilibrium. In order to remove the degeneracy one should fix the direction  of the unit vector $\vec{e}$. It can be
chosen to be along the $z$ direction. Then all the quasiaverages will be the definite numbers. This is the kind 
that one usually deals with in the theory of ferromagnetism.\\
The value of the quasi-average (\ref{q9}) may depend on the concrete structure of the additional 
term  $\Delta H = H_{\nu} - H$, if the dynamical 
quantity to be averaged is not invariant with respect to the symmetry group of the original Hamiltonian $H$. For a degenerate 
state the limit of ordinary averages (\ref{q10}) as the inclusion parameters $\nu$  of the sources tend to zero in an arbitrary 
fashion, may not exist. For a complete definition of quasiaverages it is necessary to indicate the manner in which these 
parameters tend to zero in order to ensure convergence~\cite{nnbj}. On the other hand, in order to remove degeneracy it 
suffices, in the construction of $H$, to violate only those additive 
conservation laws whose  switching  lead  to instability of the ordinary average. 
Thus  in terms of quasiaverages the selection rules for 
the correlation functions~\cite{nnb61w,petr95} that are not relevant are those that are restricted by these conservation 
laws.\\
By using $H_{\nu}$, we define the state $\omega(A) = \langle A \rangle_{\nu}$ and then let $\nu$ tend to 
zero (after passing to the thermodynamic limit)~\cite{nnb61,nnb61w,petr95}. If all averages $\omega(A)$ get infinitely small increments under 
infinitely small perturbations $\nu$, this means that
the state of statistical equilibrium under consideration is nondegenerate~\cite{nnb61,nnb61w,petr95}. However, if some states 
have finite increments
as $\nu \rightarrow 0$, then the state is degenerate. In this case, instead of ordinary averages $\langle A \rangle_{H}$, one should introduce 
the quasiaverages (\ref{q9}), for which the usual selection rules do not hold.\\
The method of quasiaverages is directly related to the principle  weakening of the  correlation~\cite{nnb61,nnb61w,petr95} in 
many-particle systems. According to this principle, the notion of the weakening of the  correlation, known in
statistical mechanics~\cite{nnb61,nnb61w,petr95}, in the case of state degeneracy must be interpreted in the sense 
of the quasiaverages~\cite{nnb61w}.\\
The quasiaverages may be obtained from the ordinary averages by using the cluster property which was formulated
by Bogoliubov~\cite{nnb61w}. This was first done when deriving the Boltzmann equations from the chain of equations
for distribution functions, and in the investigation of the model Hamiltonian in the theory 
of superconductivity~\cite{nnb60,nnb61,bb09,bs75,petr95}.  To demonstrate this
let us   consider averages (quasiaverages) of the form
\begin{equation}
\label{q11}
F(t_{1},x_{1}, \ldots t_{n},x_{n}) = \langle \ldots \Psi^{\dag}(t_{1},x_{1}) \ldots \Psi(t_{j},x_{j}) \ldots \rangle,
\end{equation}
where the number of creation operators $\Psi^{\dag}$ may be not equal to the number of annihilation operators $\Psi$. We 
fix times and split the arguments $(t_{1},x_{1}, \ldots t_{n},x_{n})$ into several 
clusters $( \ldots , t_{\alpha},x_{\alpha}, \ldots ), \ldots ,$
$( \ldots , t_{\beta},x_{\beta}, \ldots ).$  Then it is reasonably to assume that the
distances between all clusters $|x_{\alpha} - x_{\beta}|$ tend to infinity. Then, according to the cluster property, the
average value (\ref{q11}) tends to the product of averages of collections of operators with the arguments
$( \ldots , t_{\alpha},x_{\alpha}, \ldots ), \ldots ,$  $( \ldots , t_{\beta},x_{\beta}, \ldots )$ 
\begin{equation}
\label{q12}
\lim_{|x_{\alpha} - x_{\beta}| \rightarrow \infty}  F(t_{1},x_{1}, \ldots t_{n},x_{n}) = 
F( \ldots , t_{\alpha},x_{\alpha}, \ldots )  \ldots   F( \ldots , t_{\beta},x_{\beta}, \ldots ).
\end{equation}
For equilibrium states with small densities and short-range potential, the validity of this property can be 
proved~\cite{petr95}.
 For the general case, the validity of the cluster property has not yet been proved. Bogoliubov
formulated it not only for ordinary averages but also for quasiaverages, i.e., for anomalous averages, too. It works
for many important models, including the models of superfluidity~\cite{bb09,petr95} and 
superconductivity~\cite{bb09,petr95,nnb58,bts58} ( see also Refs.~\cite{pit03,peth02,grif09}).\\
To illustrate  this statement consider Bogoliubov's theory of a Bose-system with separated condensate, which is given by the 
Hamiltonian~\cite{bb09,petr95}
\begin{eqnarray}\label{q13}
H_{\Lambda}  = \int_{\Lambda}\Psi^{\dag}(x)(- \frac{\Delta}{2m})\Psi(x)dx - \mu \int_{\Lambda}\Psi^{\dag}(x)\Psi(x)dx \\ \nonumber
+ \frac{1}{2} \int_{\Lambda^{2}}\Psi^{\dag}(x_{1}) \Psi^{\dag}(x_{2}) \Phi (x_{1} - x_{2})   \Psi(x_{2})\Psi(x_{1})dx_{1} dx_{2}.
\end{eqnarray} 
This Hamiltonian can be written also in the following form
\begin{eqnarray}\label{q14}
H_{\Lambda}  =  H_{0} + H_{1} = \int_{\Lambda}\Psi^{\dag}(q)(- \frac{\Delta}{2m})\Psi(q)d q \\ \nonumber
+ \frac{1}{2} \int_{\Lambda^{2}}\Psi^{\dag}(q) \Psi^{\dag}(q') \Phi (q - q')   \Psi(q')\Psi(q)d q d q'.
\end{eqnarray} 
Here, $\Psi(q)$, and $\Psi^{\dag}(q)$  are the operators of annihilation and creation of bosons. They satisfy the 
canonical commutation relations
\begin{equation}
\label{q15}
 [\Psi(q),\Psi^{\dag}(q')]  = \delta (q - q'); \quad [\Psi(q),\Psi(q')]  = [\Psi^{\dag}(q),\Psi^{\dag}(q')]  = 0.
\end{equation}
The system of bosons is contained in the cube $A$ with the edge $L$ and volume $V$. It was assumed that it satisfies
periodic boundary conditions and the potential $\Phi(q)$ is spherically symmetric and proportional 
to the small parameter.
It was also assumed that, at temperature zero, a certain macroscopic number of particles having a nonzero
density is situated in the state with momentum zero. 
The operators $\Psi(q)$, and $\Psi^{\dag}(q)$  can be represented in the form
\begin{equation}
\label{q16}
  \Psi(q)  = a_{0}/\sqrt{V};  \quad \Psi^{\dag}(q) = a^{\dag}_{0}/\sqrt{V},
\end{equation}
where $a_{0}$ and $a^{\dag}_{0}$ are the operators of annihilation and creation of particles with momentum zero.\\
To explain the phenomenon of
superfluidity, one should calculate the spectrum of the Hamiltonian, which is quite a difficult problem. Bogoliubov
suggested the idea of approximate calculation of the spectrum of the ground state and its elementary excitations
based on the physical nature of superfluidity. His idea consists of a few assumptions. The main assumption
 is   that at temperature zero the macroscopic number of particles (with nonzero density) has the
momentum zero. Therefore, in the thermodynamic limit, the operators $a_{0}/\sqrt{V}$ and $a^{\dag}_{0}/\sqrt{V}$ commute
\begin{equation}
\label{q17}
\lim_{V \rightarrow \infty} \left [ a_{0}/\sqrt{V}, a^{\dag}_{0}/\sqrt{V} \right ] = \frac{1}{V} \rightarrow 0
\end{equation}
and are $c$-numbers. Hence, the operator of the number of particles $N_{0} =a^{\dag}_{0}a_{0}$  is a $c$-number, too.
It is worth noting that the Hamiltonian (\ref{q14}) is invariant under the gauge 
transformation $\tilde{a}_{k} = \exp (i \varphi) a_{k}$, $\tilde{a}^{\dag}_{k} = \exp ( - i \varphi) a^{\dag}_{k}$, where $\varphi$ is an
arbitrary real number. Therefore, the averages $\langle a_{0}/\sqrt{V} \rangle$ and  
$\langle a^{\dag}_{0}/\sqrt{V} \rangle$ must vanish. But this contradicts to the
assumption  that $ a_{0}/\sqrt{V}$ and $a^{\dag}_{0}/\sqrt{V}$ must become $c$-numbers in the thermodynamic limit. 
In addition it must be taken into account that $a^{\dag}_{0}a_{0}/V = N_{0}/V \neq 0$
which implies that $a_{0}/\sqrt{V} = N_{0} \exp (i \alpha)/\sqrt{V} \neq 0$ and 
$a^{\dag}_{0}/\sqrt{V} = N_{0}\exp (- i \alpha)/\sqrt{V} \neq 0$, where $\alpha$ is an arbitrary
real number. This contradiction may be overcome if we assume that the eigenstates of the Hamiltonian are degenerate
and not invariant under gauge transformations, i.e., that a spontaneous breaking of symmetry takes place.\\
Thus the averages $\langle a_{0}/\sqrt{V} \rangle$ and $\langle a^{\dag}_{0}/\sqrt{V} \rangle$, which are nonzero 
under spontaneously broken gauge invariance,
are called anomalous averages or \emph{quasiaverages}. This innovative idea of Bogoliubov penetrate deeply into the 
modern quantum physics. The systems with spontaneously
broken symmetry are studied by use of the transformation of the operators of the form 
\begin{equation}
\label{q18}
 \Psi(q)  = a_{0}/\sqrt{V} + \theta(q);  \quad \Psi^{\dag}(q) = a^{\dag}_{0}/\sqrt{V} + \theta^{*}(q),
\end{equation}
where $a_{0}/\sqrt{V}$ and $a^{\dag}_{0}/\sqrt{V}$ are the numbers first introduced by Bogoliubov in 1947 in his investigation of the
phenomenon of superfluidity~\cite{nnb61,bb09,petr95}. The main conclusion was made that for the systems with spontaneously broken symmetry, the 
quasiaverages should be studied instead of the ordinary averages. It turns out that the long-range order appears not only in the system of
Bose-particles but also in all systems with spontaneously broken symmetry. Bogoliubov's papers outlined above
anticipated the methods of investigation of systems with spontaneously broken symmetry for many years.\\ 
As mentioned above, in order to explain the phenomenon of superfluidity, Bogoliubov assumed that the operators
$a_{0}/\sqrt{V}$ and $a^{\dag}_{0}/\sqrt{V}$ become $c$-numbers in the thermodynamic limit. This statement was rigorously 
proved in the papers by Bogolyubov and some other authors.
Bogolyubov's proof was based on the study of the equations for two-time Green's functions~\cite{bt59} and on the assumption
that the cluster property holds. It was proved that the solutions of equations for Green's functions for the system
with Hamiltonian (\ref{q14})  coincide with the solutions of the equations for the system with the same Hamiltonian in
which the operators $a_{0}/\sqrt{V}$ and $a^{\dag}_{0}/\sqrt{V}$ are replaced by numbers. These numbers should be determined from
the condition of minimum for free energy. Since all the averages in both systems coincide, their free energies coincide, too.\\
It is worth noting that the validity of the replacement of the operators 
$a_{0}$ and $a^{\dag}_{0}$ by $c$-numbers in the thermodynamic limit was confirmed  in the
numerous subsequent publications of various authors. 
Thus Bogoliubov's 1947 analysis of the many-body Hamiltonian by means of a
$c$-number substitution for the most relevant operators in
the problem, the zero-momentum mode operators, was justified rigorously.
Since the Bogoliubov's 1947 analysis is one of the key developments in the theory of the Bose
gas, especially the theory of the low density gases currently
at the forefront of experiment~\cite{pit03,peth02,grif09}, this result is of importance for the legitimation of that theory. Additional arguments
were given in study, where the 
Bose-Einstein condensation and spontaneous $U(1)$ symmetry breaking were investigated on the
basis of Bogoliubov's truncated Hamiltonian $H_{B}$ for a weakly interacting Bose system, and adding a $U(1)$
symmetry breaking term $ \sqrt{V} ( \lambda a_{0} +  \lambda^{*} a^{\dag}_{0})$
 to $H_{B}$, 
It was shown also, by using the coherent state theory and the mean-field
approximation rather than the $c$-number approximations, that the Bose-Einstein condensation   occurs if and only
if the $U(1)$ symmetry of the system is spontaneously broken. The real ground state energy and the justification of the
Bogoliubov $c$-number substitution were given by solving the Schroedinger eigenvalue equation and using the self-consistent
condition. Thus the Bogoliubov
$c$-number substitutions were fully correct and   the symmetry breaking causes the displacement of the condensate
state.\\
The concept of quasiaverages was introduced by Bogoliubov on the basis of an analysis of many-particle systems with
a degenerate statistical equilibrium state. Such states are inherent to various physical many-particle systems~\cite{bb09,petr95}. Those are
liquid helium in the superfluid phase, metals in the superconducting state, magnets in the ferromagnetically ordered state,
liquid crystal states, the states of superfluid nuclear matter, etc. (for a review, see Refs.~\cite{kuz09,kov06}).
In case of superconductivity, the 
source 
$\nu \sum_{k}v(k)(a^{\dag}_{k\uparrow}a^{\dag}_{-k\downarrow} + a_{-k\downarrow}a_{k\uparrow})$ 
was inserted 
in the BCS-Bogoliubov Hamiltonian, and the quasiaverages were defined by use of the Hamiltonian $H_{\nu}$. 
In the general case, the sources are introduced to remove degeneracy. If infinitesimal sources give infinitely small contributions
to the averages, then this means that the corresponding degeneracy is absent, and there is no reason to insert
sources in the Hamiltonian. Otherwise, the degeneracy takes place, and it is removed by the sources. The ordinary
averages can be obtained from quasiaverages by averaging with respect to the parameters that characterize the
degeneracy.\\
N. N. Bogoliubov, Jr.~\cite{nnbj} considered some features of quasiaverages for model systems with four-fermion interaction.
He discussed the treatment of certain three-dimensional model systems which can be solved exactly. For this aim
a new effective way of defining quasiaverages for the systems under consideration was proposed.\\
Peletminskii and Sokolovskii~\cite{peso74} have found general expressions for the operators of the flux densities of physical
variables in terms of the density operators of these variables. The method of quasiaverages and the expressions 
found for the flux operators were used to obtain the averages of these operators in terms of the thermodynamic
potential in a state of statistical equilibrium of a superfluid liquid. \\ Vozyakov~\cite{voz79} reformulated the theory
of quantum crystals in terms of quasiaverages. He analyzed a Bose system with periodic distribution of particles
which simulates an ensemble in which the particles cannot be regarded as vibrating independently about a position of 
equilibrium lattice sites. With allowance for macroscopic filling of the states corresponding to the
distinguished symmetry, a calculation was made of an excitation spectrum in which there exists a collective
branch of gapless type.\\
Peregoudov~\cite{pere97} discussed
the effective potential method, used in quantum field theory to study spontaneous symmetry breakdown,  
 from the point of view of Bogoliubov's quasiaveraging procedure. It was shown that the effective
potential method is a disguised type of this procedure. The catastrophe theory approach to the study of
phase transitions was discussed and the existence of the potentials used in that approach was proved from the
statistical point of view. It was shown that in the ease of broken symmetry, the nonconvex effective potential
is not a Legendre transform of the generating functional for connected Green's functions. Instead, it is a
part of the potential used in catastrophe theory. The relationship between the effective potential and the
Legendre transform of the generating functional for connected Green's functions is given by  Maxwell's rule.
A rigorous rule for evaluating quasiaveraged quantities within the framework of the effective  potential
method was established.\\
N. N. Bogoliubov, Jr. with M. Yu. Kovalevsky and co-authors~\cite{kov09} developed a statistical approach for solving 
the problem of classification of equilibrium states in condensed media with spontaneously broken
symmetry based on the quasiaverage concept. 
Classification of equilibrium states of condensed media with spontaneously broken symmetry was
carried out. The generators of residual and spatial symmetries
were introduced and equations of classification for the order
parameter has been found.
Conditions of residual symmetry and spatial symmetry were formulated. The connection between
these symmetry conditions and equilibrium states of various media with tensor order parameter was found out.
An analytical solution of the problem of classification of
equilibrium states for superfluid media, liquid crystals and
magnets with tensor order parameters was obtained.
Superfluid $^{3}He$, liquid crystals, quadrupolar magnetics were considered in detail. Possible homogeneous and
heterogeneous states were found out. Discrete and continuous thermodynamic parameters, which define an
equilibrium state, allowable form of order parameter, residual symmetry, and spatial symmetry generators
were established.
This approach, which is alternative to the well-known Ginzburg-Landau method, 
does not contain any model assumptions concerning the form of the free energy as functional of
the order parameter and does not employ the requirement of
temperature closeness to the point of phase transition. For all
investigated cases they found the structure of the order parameters and the explicit forms of generators of residual and
spatial symmetries.
Under the certain restrictions they established the form of
the order parameters in case of spins $0$, $1/2$, $1$ and proposed
the physical interpretation of the studied degenerate states of
condensed media. 
%
%
%
%
\section{  Quantum Protectorate}
%
%
%
The "quantum protectorate" concept was formulated
in  paper~\cite{pnas}. Its inventors, R. Laughlin and
D. Pines, discussed the most fundamental principles of
matter description in the widest sense of this word.
They formulated their main thesis: emergent physical
phenomena, which are regulated by higher physical
principles, have a certain property, typical for these
phenomena only. This property is their insensitivity to
microscopic description. For instance,
the crystalline state is the simplest known example of a
quantum protectorate, a stable state of matter whose generic
low-energy properties are determined by a higher organizing
principle and nothing else. There are many other examples~\cite{pnas}.
These quantum protectorates,
with their associated emergent behavior, provide us
with explicit demonstrations that the underlying microscopic
theory can easily have no measurable consequences whatsoever
at low energies. \textbf{The nature of the underlying theory is unknowable
until one raises the energy scale sufficiently to escape
protection.} The existence of two scales, the low-energy and
high-energy scales, relevant to the description of magnetic
phenomena was stressed by the author of the present
work in the papers~\cite{kuz09,kuz00,kuz02}, which were devoted to comparative
analysis  of localized and band models of
quantum theory of magnetism. It was shown there, that
the low-energy spectrum of magnetic excitations in the
magnetically-ordered solid bodies corresponds to a
hydrodynamic pole ($\vec{k}, \omega \rightarrow 0 $) in the generalized
spin susceptibility  $\chi$, which is present in the Heisenberg,
Hubbard, and the combined $s-d$ model. In
the Stoner band model the hydrodynamic pole is
absent, there are no spin waves there. At the same time,
the Stoner single-particle's excitations are absent in the
Heisenberg model's spectrum. The Hubbard model  
with narrow energy bands contains both types
of excitations: the collective spin waves (the low-energy
spectrum) and Stoner single-particle's excitations
(the high-energy spectrum). This is a big advantage
and flexibility of the Hubbard model in comparison
to the Heisenberg model. The latter, nevertheless, is
a very good approximation to the realistic behavior in
the limit $\vec{k}, \omega \rightarrow 0,$
the domain where the hydrodynamic description is
applicable, that is, for long wavelengths and low energies.
The quantum protectorate concept was applied to
the quantum theory of magnetism by the the present author  
in  paper~\cite{kuz02}, where a criterion of applicability of models
of the quantum theory of magnetism  to
description of concrete substances was formulated. The
criterion is based on the analysis of the model's low-energy
and high-energy spectra. 
 There the detailed analysis was carried out
of the idea of quantum protectorate~\cite{pnas}  in the
context of quantum theory of magnetism~\cite{kuz02}.  It was suggested that the
difficulties in the formulation of   quantum theory of magnetism
at the  microscopic level,  that are related to the choice of
relevant models,  can be understood better in the light of the quantum protectorate
concept~\cite{kuz02}. We argued that the difficulties in the formulation of
adequate microscopic models of electron and magnetic properties of
materials are intimately related to dual, \textbf{itinerant} and \textbf{localized}
behavior of electrons~\cite{kuz00}. We formulated a criterion of what basic
picture describes best  this dual behavior. The main suggestion
was that quasiparticle excitation spectra might provide
distinctive signatures and good criteria for the appropriate
choice of the relevant model.
A broad class of the problems of condensed matter physics~\cite{kuz09,kuz10} in the fields of  the 
magnetism and superconductivity of  
complex materials were reconsidered in relation to these ideas. 
%
%
%
\section{Irreducible Green Functions Method} 
%
%
It was shown above that it becomes clear that only a thorough experimental and theoretical investigation 
of quasiparticle many-body dynamics
of the many particle systems can provide the answer on the relevant microscopic picture.
Many-particle systems where the interaction 
is strong have often complicated behavior, and require nonperturbative approaches to treat their properties.
There are many different approaches to construction
of generalized mean-field approximations; however, all
of them have a special-case character. The method of
irreducible Green  functions (IGF)  allows one to tackle this
problem in a more systematic fashion. In order to clarify
this statement let us consider briefly
the main ideas of the IGF approach that allows one to describe
completely  quasiparticle spectra with damping in a very natural way.
When working with infinite hierarchies of equations
for Green  functions the main problem is finding the methods for
their efficient decoupling, with the aim of obtaining a
closed system of equations, which determine the Green  functions. A
decoupling approximation must be chosen individually
for every particular problem, taking into account its
character. This "individual approach" is the source of
critique for being too not transparent, which sometimes
appear in the papers using the causal Green  functions and diagram
technique. However, the ambiguities are also present in
the diagram technique, when the choice of an appropriate
approximation is made there. The decision, which
diagrams one has to sum up, is obvious only for a narrow
range of relatively simple problems. In the paper~\cite{plak73}
 devoted to Bose-systems, and in the papers
by the author of present work~\cite{kuz78,kuz89,rnc02}
 devoted to Fermi systems it was shown that for a
wide range of problems in statistical mechanics and
theory of condensed matter one can outline a fairly systematic
recipe for constructing approximate solutions
in the framework of irreducible Green's functions
method. Within this approach one can look from a unified
point of view at the main problems of fundamental
characters arising in the method of two-time temperature
Green  functions. The method of irreducible Green  functions is
a useful reformulation of the ordinary Bogoliubov-Tyablikov method of equations of motion~\cite{tyab,bt59}.\\
We  reformulated  the two-time Green  functions method~\cite{tyab,kuz78,kuz89,rnc02} to the
form which is especially adjusted  to correlated fermion systems
on a lattice and systems with complex spectra. A very important
concept of the whole method is the {\it generalized mean fields}
(GMFs), as it was formulated in~\cite{kuz09,rnc02}. These GMFs have a
complicated structure for a strongly correlated case and complex
spectra, and are not reduced to the functional of  mean densities
of the electrons or spins when
one calculates excitation spectra at finite temperatures. \\
To clarify the foregoing, let us consider a retarded Green  function of the
form~\cite{tyab}
\begin{equation}
\label{eq6} G^{r} = \langle \langle A(t), A^{\dagger}(t')\rangle \rangle = -i\theta(t -
t')\langle [A(t) A^{\dagger}(t')]_{\eta}\rangle, \eta = \pm 1.
\end{equation}
As an introduction to the concept of IGF, let us describe the
main ideas of this approach in a symbolic and simplified form. To
calculate the retarded Green  function $G(t - t')$,  let us write down the
equation of motion for it
\begin{equation}
\label{eq7} \omega G(\omega) = \langle [A, A^{\dagger}]_{\eta} \rangle + \langle \langle [A,
H]_{-}\mid A^{\dagger} \rangle \rangle_{\omega}.
\end{equation}
Here we use the notation $ \langle \langle A(t), A^{\dagger}(t') \rangle \rangle$ for the time-dependent Green  function and
$\langle \langle A \mid A^{\dagger} \rangle \rangle_{\omega}$ for its Fourier transform~\cite{tyab}. The notation $[A,B]_{\eta}$
refers to commutation and anticommutation, depending on the value of $\eta = \pm$. 
The essence of the method
is as follows~\cite{rnc02}.   It is based on the notion of the
{\it "IRREDUCIBLE"} parts of Green  functions (or the irreducible parts of the
operators, $A$ and $A^{\dagger}$, out of which the Green  function is
constructed) in terms of which it is possible, without recourse
to a truncation of the hierarchy of equations for the Green  functions, to
write down the exact Dyson equation and to obtain an exact
analytic representation for the self-energy operator. By
definition, we introduce the irreducible part {\bf (ir)} of the Green  function 
\begin{equation}
\label{eq8} ^{(ir)}\langle \langle [A, H]_{-}\vert A^{\dagger}\rangle \rangle = \langle \langle [A, H]_{-}
- zA\vert A^{\dagger} \rangle \rangle.
\end{equation}
The unknown constant z is defined by the condition (or constraint)
\begin{equation}
\label{eq9} \langle [[A, H]^{(ir)}_{-}, A^{\dagger}]_{\eta} \rangle = 0,
\end{equation}
which is an analogue of the orthogonality condition in the Mori
formalism~\cite{lee1,lee2}. 
Let us emphasize that due to the complete equivalence of the definition of the
irreducible parts for the Green  functions $(^{(ir)}\langle \langle [A, H]_{-}\vert A^{\dagger} \rangle \rangle)$ 
and operators  $(^{(ir)}[A, H]_{-}) \equiv ([A, H]_{-})^{(ir)}$ we will use both the notation 
freely  ( $ ^{(ir)}\langle \langle A  \vert B \rangle \rangle$  is the same as
$\langle \langle (A)^{(ir)}  \vert B \rangle \rangle$ ).  A choice  
one notation over another is determined by the brevity and clarity of notation only.
From the condition (\ref{eq9}) one can find
\begin{equation}
\label{eq10} z = \frac{\langle [[A, H]_{-}, A^{\dagger}]_{\eta} \rangle}{\langle [A,
A^{\dagger}]_{\eta} \rangle} =
 \frac{M_{1}}{M_{0}}.
\end{equation}
Here $M_{0}$ and $M_{1}$ are the zeroth and first order moments
of the spectral density. Therefore, the irreducible Green  functions  are
defined so that they cannot be reduced to the lower-order ones by
any kind of decoupling. It is worth  noting that the term {\it
"irreducible"} in a group theory means a representation of a
symmetry operation that cannot be expressed in terms of lower
dimensional representations. Irreducible (or connected )
correlation functions are known in statistical mechanics. In the
diagrammatic approach, the irreducible vertices are defined as
graphs that do not contain inner parts connected by the
$G^{0}$-line. With the aid of the definition (\ref{eq8})   these
concepts are expressed in terms  of retarded and advanced Green  functions.
The procedure extracts all relevant (for the problem under
consideration) mean-field contributions and puts them into the
generalized mean-field Green  function which  is defined here as
\begin{equation}
\label{eq11} G^{0}(\omega) = \frac{\langle [A,
A^{\dagger}]_{\eta} \rangle}{(\omega - z)}.
\end{equation}
To calculate the IGF $\,  ^{(ir)}\langle \langle [A, H]_{-}(t),
A^{\dagger}(t') \rangle \rangle$ in (\ref{eq7}), we have to write the equation
of motion for it after differentiation with respect to the second
time variable $t'$. The condition of orthogonality (\ref{eq9})
removes the inhomogeneous term from this equation and is a very
crucial point of the whole approach. If one introduces the
irreducible part for the right-hand side operator, as discussed
above for the ``left" operator, the equation of motion
(\ref{eq7}) can be exactly rewritten in the following form:
\begin{equation} \label{eq12} G = G^{0} + G^{0}PG^{0}.
\end{equation}
The scattering operator $P$ is given by
\begin{equation}
\label{eq13} P = (M_{0})^{-1}(\, ^{(ir)}\langle \langle [A,
H]_{-}\vert[A^{\dagger}, H]_{-} \rangle \rangle^{(ir)}) (M_{0})^{-1}.
\end{equation}
The structure of  equation ( \ref{eq13}) enables us to determine
the self-energy operator $M$  by  analogy with the diagram
technique
\begin{equation} \label{eq14} P = M + MG^{0}P.
\end{equation}
We used here the notation $ M $ for self-energy (mass operator
in  quantum field theory).
From the definition (\ref{eq14}) it follows that  the self-energy
operator $M$ is defined as a proper (in the diagrammatic language,
``connected") part of the scattering operator $M = (P)^{p}$. As a
result, we obtain the exact Dyson equation for the thermodynamic
double-time Green functions
\begin{equation} \label{eq15} G =
G^{0} + G^{0} M G.
\end{equation}
The difference between $P$ and $M$ can be regarded as two
different solutions of two integral equations (\ref{eq12}) and
(\ref{eq15}). However, from  the Dyson equation (\ref{eq15}) only
the full GF  is seen to be expressed as a  formal solution of the
form
\begin{equation}
\label{eq16} G = [ (G^{0})^{-1} - M ]^{-1}.
\end{equation}
Equation  (\ref{eq16}) can be regarded as an alternative form of
the Dyson equation (\ref{eq15}) and the {\it definition} of $M$
provides that the generalized mean-field GF $G^{0}$ is specified.
On the contrary , for the scattering operator $P$, instead of the
property $G^{0}G^{-1} + G^{0}M = 1$, one has the property
$$(G^{0})^{-1} - G^{-1} = P G^{0}G^{-1}.$$  Thus, the { \it very functional
form} of the formal solution (\ref{eq16})   precisely determines
the difference between $P$ and $M$. \\ Thus, by introducing
irreducible parts of GF (or  irreducible parts of the operators,
out of which the GF is constructed) the equation of motion
(\ref{eq7}) for the GF can exactly be  (but using the
orthogonality constraint (\ref{eq9})) transformed into the Dyson
equation for the double-time thermal GF (\ref{eq15}). This result
is very remarkable  because  the traditional form of the GF
method does not include this point. Notice that all quantities
thus considered are  exact. Approximations can be generated not
by truncating the set of coupled equations of motions but by a
specific approximation of the functional form of the mass
operator $M$ within a self-consistent scheme  expressing $M$ in
terms of the initial GF
$$ M \approx F[G].$$
Different approximations are relevant to different physical
situations. The projection operator technique  has essentially
the same philosophy. But with using the constraint (\ref{eq9}) in
our approach we emphasize the fundamental and central role of the
Dyson equation for  calculation of single-particle properties of
many-body systems. The problem of reducing the whole hierarchy of
equations involving higher-order GFs by a coupled nonlinear set
of integro-differential equations connecting the single-particle
GF to the self-energy operator is rather nontrivial. A
characteristic feature of these equations is that  besides the
single-particle GF  they involve also higher-order GF. The
irreducible counterparts of the GFs, vertex functions, serve to
identify correctly the self-energy as
$$  M = G^{-1}_{0}  - G^{-1}.$$
The integral form of the Dyson equation (\ref{eq15}) gives  $M$
the physical meaning of a nonlocal and energy-dependent effective
single-particle potential. This meaning can be verified for the
exact self-energy using the diagrammatic expansion for the
causal GF.\\
It is important to note that for the retarded and advanced GFs,
the notion of the proper part $M = (P)^{p}$ is symbolic in
nature~\cite{kuz09,rnc02}. In a certain sense, it is possible to say that
it is defined here by analogy with the irreducible many-particle
$T$-matrix. Furthermore, by analogy with the diagrammatic
technique, we can also introduce the proper part defined as a
solution to the integral equation (\ref{eq14}). These analogues
allow us to better understand  the formal structure of the Dyson
equation for the double-time thermal GF, but only in a symbolic
form . However, because of the identical form of the equations
for  GFs for all three types (advanced, retarded, and causal),
we can convert our calculations  to causal GF  at each stage of
calculations  and, thereby, confirm the substantiated nature of
definition (\ref{eq14}). We therefore should speak of an analogy
of the Dyson equation. Hereafter, we  drop this stipulating,
since it does not cause any misunderstanding. In a sense, the IGF
method is a variant of the Gram-Schmidt orthogonalization
procedure~\cite{rnc02}.\\ It should be emphasized that the scheme presented
above gives just a general idea of the IGF method. A more exact
explanation why one should not introduce the approximation
already in $P$, instead of having to work out $M$, is given below
when working out the application
of the method to  specific problems.\\
The general philosophy of the IGF method is in the separation and
identification of elastic scattering effects and inelastic ones.
This latter point is quite often underestimated, and both effects
are mixed. However, as far as the right definition of
quasiparticle damping is concerned, the separation of elastic
and inelastic scattering processes is believed to be crucially
important for  many-body systems with complicated spectra and
strong interaction.   \\ From a technical point of view, the
elastic GMF renormalizations can exhibit  quite a nontrivial
structure. To obtain this structure correctly, one should
construct the full GF from the complete algebra of  relevant
operators and develop a special projection procedure for
higher-order GFs, in accordance with a given algebra. Then a
natural question arises how to select the relevant set of
operators $\{ A_{1}, A_{2}, ... A_{n} \}$   describing the
"relevant degrees of freedom". The above consideration suggests
an intuitive and heuristic way to the suitable procedure as
arising from an infinite chain of equations of motion
(\ref{eq7}). Let us consider the column
$$ \begin{pmatrix}
 A_{1}\cr  A_{2}\cr \vdots \cr  A_{n}\cr
\end{pmatrix}, $$
where
$$ A_{1} = A,\quad A_{2} = [A,H],\quad A_{3} = [[A,H],H], \ldots
A_{n} = [[... [A, \underbrace{H]...H}_{n}].$$
Then the most
general possible Green function can be expressed as a matrix
$$ \hat G = \langle \langle \begin{pmatrix}
A_{1}\cr  A_{2}\cr \vdots \cr A_{n}\cr \end{pmatrix} \vert  \begin{pmatrix}
A^{\dagger}_{1}& A^{\dagger} _{2}& \ldots &  A^{\dagger}
_{n}\cr \end{pmatrix} \rangle \rangle $$
This generalized Green function describes the one-, two-, and
$n$-particle dynamics. The equation of motion for it includes, as
a particular case, the Dyson equation for single-particle Green
function, and the Bethe-Salpeter equation which is the equation
of motion for the two-particle Green function and which is an
analogue of the Dyson equation, etc . The corresponding reduced
equations should be extracted from the equation of motion for the
generalized Green function with the aid of special techniques such as the
projection method and similar techniques. This must be a final
goal towards a real understanding of the true many-body dynamics.
At this point, it is worthwhile to underline that the above
discussion is a heuristic scheme only,  but not a straightforward
recipe. The specific method of introducing  the IGFs depends on
the form of operators $A_{n}$, the type of the Hamiltonian, and
conditions of
the problem. 
Here a sketchy form of the IGF method was presented. The aim was to
introduce the general scheme and to lay the groundwork for
generalizations.  We  demonstrated  in~\cite{kuz09,rnc02} that the
IGF method is a powerful tool for describing the quasiparticle
excitation spectra, allowing a deeper understanding of elastic and
inelastic quasiparticle scattering effects and the corresponding
aspects of damping and finite lifetimes. In a certain sense,
it provides a clear link between the equation-of-motion approach
and the diagrammatic methods due to derivation of the Dyson
equation.  Moreover, due to the fact that it allows the
approximate treatment of the self-energy effects on a final stage,
it yields a systematic way of the construction of approximate
solutions.
%
\section{Effective and Generalized Mean Fields}
%
%
The most common technique for studying the subject of interacting
many-particle systems is to use the mean field theory. This
approximation was especially popular in the theory of
magnetism~\cite{tyab,kuz09,rnc02}. 
To calculate the susceptibility and other characteristic
functions of a system of localized magnetic moments, with a given
interaction Hamiltonian, the approximation, termed the "molecular
field approximation" was  used widely. However, it is not an easy
task to give the formal unified definition what the mean field
is. In a sense, the mean field is the umbrella term for a variety
of theoretical methods of reducing  the many-particle problem to
the single-particle one. Mean field theory, that approximates the
behavior of a system by ignoring the effect of fluctuations and
those spin correlations which dominate the collective properties
of the ferromagnet usually provides a starting and estimating
point only, for studying phase transitions. The mean field
theories miss important features of the dynamics of a system. The
main intention of the mean field theories, starting from the
works of J. D. van der Waals and P. Weiss, is to take into account the
cooperative behavior of a large number of particles. It is well
known that earlier theories of phase transitions based on the
ideas of van der Waals and Weiss lead to predictions which are
qualitatively at variance with results of measurements near the
critical point. Other variants of simplified mean field theories
such as the Hartree-Fock theory for electrons in atoms,   
lead to discrepancies of various kinds too. It is therefore
natural to analyze the reasons for such drawbacks of earlier
variants of the mean field theories.\\
A number of effective field theories which are improved versions
of the "molecular field approximation" were proposed. In our papers~\cite{kuz09,kuz78,kuz89,rnc02}
we stressed a specificity of strongly
correlated many-particle systems on a lattice contrary to
continuum (uniform) systems.  The earlier concepts of molecular field were described
in terms of a functional of mean magnetic moments (in magnetic
terminology) or mean particle densities.
The corresponding mean-field functional $F[\langle n \rangle, \langle S^{z} \rangle]$
describes the {\it uniform} mean field.  Actually, the Weiss
model was not based on discrete "spins"  as is well known, but
the uniformity of the mean internal field was the most essential
feature of the model. In the modern language, one should assume
that the interaction between atomic spins $S_{i}$ and its
neighbors is equivalent to a mean (or molecular) field, $ M_{i}
= \chi_{0} [ h_{i}^{(ext)} + h_{i}^{(mf)} ] $ and that the
molecular field $ h_{i}^{(mf)} $ is of the form $ h^{(mf)} =
\sum_{i} J(R_{ji})\langle S_{i} \rangle$ (above $T_{c}$ ). Here $ h^{ext}$ is
an applied conjugate field, $\chi_{0}$ is the response function,
and $ J(R_{ji})$ is an interaction. In  other words, the mean
field approximation reduces the many-particle problem to a
single-site  problem in which a magnetic moment at any site can
be either parallel or antiparallel to the total magnetic field
composed of the applied field and the molecular field. The
average interaction of  $i$ neighbors was taken into account only,
and the fluctuations were neglected. One particular example,
where the mean field theory works relatively well is the
homogeneous structural phase transitions; in this case the
fluctuations are confined in phase space.  The next important
step was made by L. Neel. He conjectured that the
Weiss internal field might be either positive or negative in
sign. In the latter case, he showed that below a critical
temperature (Neel temperature) an ordered arrangement of equal
numbers of oppositely directed atomic moments could be
energetically favorable. This new magnetic structure was termed
antiferromagnetism. It was conjectured that  the two-sublattice
Neel (classical) ground
state is formed by  local staggered internal mean fields. \\
There is a number of the "correlated effective field" theories,
that tend to repair the limitations of  simplified mean field
theories. The remarkable and ingenious one is the Onsager
"reaction field approximation". He suggested that the
part of the molecular field on a given dipole moment which comes
from the reaction of  neighboring molecules to the instantaneous
orientation of the moment should not be included into the
effective orienting field. This "reaction field" simply follows
the motion of the moment and thus does not favor one orientation
over another (for details see Refs.~\cite{kuz09,rnc02}).\\ 
It is known~\cite{sold}  that mean-field
approximations, for example the molecular field approximation for
a spin system, the Hartree-Fock approximation and the
BCS-Bogoliubov approximation for an electron system are
universally formulated by the Bogoliubov
inequality:
\begin{eqnarray}
\nonumber
   - \beta^{-1} ln (Tr e^{(-\beta H)} )\leq \\ -
\beta^{-1} ln (Tr e^{(-\beta H^{mf})}) + \frac {Tr e^{(- \beta
H^{mf})} ( H - H^{mf})}{Tr e^{(- \beta H^{mf})}}. \label{eq.63}
\end{eqnarray}
Here $F$  is the free energy, and $H^{mf}$ is a "trial" or a "mean
field" approximating Hamiltonian. This inequality gives the upper
bound of the free energy of a many-body system. It is important to
emphasize that the BCS-Bogoliubov theory of
superconductivity~\cite{nnb71,nnb58,bts58} was formulated on the
basis of a trial Hamiltonian  which  consists of a quadratic form
of creation and annihilation operators, including "anomalous" (off-diagonal) averages. 
The functional of the mean
field (for the superconducting single-band Hubbard model) is of
the following form~\cite{vkp82}:
\begin{equation} \label{eq.64}
\Sigma^{c}_{\sigma} =  U 
\begin{pmatrix}
 \langle a^{\dagger}_{i-\sigma}
a_{i-\sigma} \rangle & - \langle a_{i\sigma} a_{i-\sigma}\rangle \cr
- \langle a^{\dagger}_{i-\sigma} a^{\dagger}_{i\sigma} \rangle &
- \langle a^{\dagger}_{i\sigma} a_{i\sigma} \rangle \cr 
\end{pmatrix}.
\end{equation}
The "anomalous" off-diagonal terms fix the relevant
BCS-Bogoliubov vacuum and select the appropriate set of
solutions.  
From the point of view of
quantum many-body theory, the problem of adequate introduction of
mean fields for system of many interacting particles can be most
consistently investigated in the framework of the IGF method. A
correct calculation of the quasiparticle spectra and their
damping, particularly, for  systems with a complicated spectrum
and strong interaction~\cite{kuz09} reveals,  
 that the generalized mean fields can have very complicated
structure
which  cannot be described by a functional of the mean-particle density.\\
To illustrate the actual distinction of  description of the
generalized mean field in the equation-of-motion method for the
double-time Green functions, let us compare the two approaches,
namely, that of Tyablikov~\cite{tyab} and of Callen~\cite{cal63}.
We shall consider the Green function $\langle \langle S^{+}|S^{-} \rangle \rangle $ for the
isotropic Heisenberg model
\begin{equation}
\label{eq.65} H = - \frac{1}{2} \sum_{ij} J(i-j) \vec S_{i} \vec
S_{j}.
\end{equation}
The equation of motion  for the spin Green function
is of the form
\begin{eqnarray}
\label{eq.66}
\omega \langle \langle S_{i}^{+}|S_{j}^{-} \rangle \rangle_{\omega} = \\
\nonumber 2 \langle S^{z} \rangle\delta_{ij} + \sum_{g} J (i-g)
\langle \langle S^{+}_{i}S^{z}_{g} - S^{+}_{g}S_{i}^{z}|S_{j}^{-} \rangle \rangle_{\omega}.
\end{eqnarray}
The Tyablikov decoupling  expresses the second-order Green function in terms
of the first (initial) Green function:
\begin{equation}
\label{eq.67} \langle \langle S^{+}_{i}S_{g}^{z}|S^{-}_{j}\rangle \rangle =
\langle S^{z} \rangle \langle \langle S^{+}_{i}|S^{-}_{j} \rangle \rangle.
\end{equation}
This approximation is an RPA-type; it does not lead to the
damping of spin wave excitations
\begin{equation} \label{eq.68}
E(q) = \sum_{g} J(i-g)\langle S^{z} \rangle \exp [i(\vec R_{i} - \vec R_{g})
\vec q ] = 2\langle S^{z} \rangle(J_{0} - J_{q}).
\end{equation}
The reason for this is rather transparent. This decoupling does
not take into account  the {\it inelastic} magnon-magnon
scattering processes. In a sense, the Tyablikov approximation
consists of approximating the commutation relations of spin
operators to the extent of replacing the commutation relation
$[S^{+}_{i},S^{-}_{j}]_{-} = 2S^{z}_{i}\delta_{ij}$
by $ [S^{+}_{i},S^{-}_{j}]_{-} = 2 \langle S^{z} \rangle\delta_{ij}$ .\\
Callen~\cite{cal63} has proposed an improved decoupling
approximation in the method of Tyablikov in the following form:
\begin{equation} \label{eq.69}
\nonumber \\
\langle \langle S^{z}_{g}S_{f}^{+}|B \rangle \rangle \rightarrow \langle S^{z} \rangle \langle \langle S^{+}_{f}|B \rangle \rangle -
\alpha \langle S^{-}_{g}S^{+}_{f}\rangle \langle \langle S^{+}_{g}|B \rangle \rangle.
\end{equation}
Here $ 0 \leq \alpha \leq 1$. To clarify this point, it should be
reminded that for spin $1/2$ ( the procedure was  generalized by
Callen to an arbitrary spin), the spin operator $ S^{z}$   can be
written as $ S^{z}_{g} = S - S^{-}_{g}S^{+}_{g}$ or $S^{z}_{g} =
{1 \over 2} ( S^{+}_{g}S^{-}_{g} - S^{-}_{g}S^{+}_{g}).$  It is
easy to show that
$$
S^{z}_{g}  = \alpha S + \frac {1 - \alpha}{2} S^{+}_{g}S^{-}_{g}
- \frac {1 + \alpha}{2} S^{-}_{g}S^{+}_{g}.$$  The operator
$S^{-}_{g}S^{+}_{g}$ represents the deviation of  $\langle S^{z} \rangle$ from
$S$. In the low-temperature region, this deviation is small, and
$\alpha \sim 1$. Similarly, the operator ${1 \over 2} (
S^{+}_{g}S^{-}_{g} - S^{-}_{g}S^{+}_{g})$ represents the
deviation of $\langle S^{z} \rangle$ from 0. Thus, when $\langle S^{z} \rangle$ approaches to
zero, one can expect that $\alpha \sim 0$. Thus, in this way, it
is possible to obtain a correction to the Tyablikov decoupling
with either a positive or negative sign, or no correction at all,
or any intermediate value, depending on the choice of  $\alpha$.
The above Callen arguments are not rigorous , for, although the
difference in the operators $S^{+}S^{-}$ and $S^{-}S^{+}$ is
small if $\langle S^{z} \rangle \sim 0$, each operator makes a contribution of
the order of $S$, and it is each operator which is treated
approximately, not the difference. There are some other drawbacks
of the Callen decoupling scheme. Nevertheless, the Callen
decoupling was the first conceptual attempt to introduce the
interpolation decoupling procedure. Let us note that the choice
of $\alpha = 0$ over the entire temperature range is just the
Tyablikov decoupling (\ref{eq.67}).\\
The energy spectrum for the Callen decoupling is given by
\begin{equation} \label{eq.70}
E(q) = 2\langle S^{z} \rangle \bigl ((J_{0} - J_{q}) +\frac {\langle S^{z} \rangle}{NS^2}
\sum_{k} [ J(k) - J(k - q)] N(E(k)) \bigr ).
\end{equation}
Here $N(E(k))$ is the Bose distribution function $ N(E(k)) = [
\exp ( E(k)\beta) - 1 ]^{-1}$. This is an implicit equation for
$N(E(k))$, involving the unknown quantity $\langle S^{z} \rangle$ . For the
latter an additional equation is given~\cite{cal63}. Thus, both
these equations constitute a set of coupled equations which must
be solved self-consistently for $\langle S^{z} \rangle$.  This formulation of
the Callen decoupling scheme displays explicitly the tendency of
the improved description of the mean field. In a sense, it is
possible to say that the Callen work  dates really the idea of
the generalized mean field within the equation-of-motion method
for double-time GFs, however, in a semi-intuitive form. The next
essential steps were made by Plakida~\cite{plak73} for the
Heisenberg ferromagnet and by Kuzemsky~\cite{kuz78}
for the Hubbard model. 
Later many approximate schemes for decoupling
the hierarchy of equations for GF were proposed,
improving the Tyablikov's and Callen's decouplings. Various
approaches generalizing the random phase's
approximation in the ferromagnetism theory for wide
ranges of temperature were considered in  paper
by Czachor and Holas~\cite{hol90}.
As was mentioned above, the correct
definition of \emph{generalized mean fields} depends on the condition of
the problem, the strength of interaction, the choice of relevant
operators, and on the symmetry requirements.
 The most important conclusion to
be drawn from   the present consideration is that the GMF, in principle,
can have quite a nontrivial
structure and cannot be reduced to the mean-density functional only.
%
\section{  Quasiaverages and Irreducible Green Functions Method}
%
%
In condensed matter physics, the symmetry is important in
classifying different phases and   understanding the phase transitions between them.
There is an important distinction between the case where the broken
symmetry is continuous (e.g. translation, rotation, gauge invariance)
or discrete (e.g. inversion, time reversal symmetry)~\cite{kuz09}. The Goldstone theorem
states that when a continuous symmetry is spontaneously broken
and the interactions are short ranged a collective mode (excitation) exists
with a gapless energy spectrum (i.e. the energy dispersion curve
starts at zero energy and is continuous). Acoustical phonons in a crystal
are prime examples of such so-called gapless Goldstone modes. Other
examples are the Bogoliubov sound modes in (charge neutral) Bose condensates~\cite{pit03,peth02,grif09}
and spin waves (magnons) in ferro- and antiferromagnets. 
N. N. Bogoliubov and then Y. Nambu in in their works  shown that the general features of superconductivity 
are in fact model independent consequences of the spontaneous breakdown of electromagnetic gauge invariance. 
It is important to emphasize that the BCS-Bogoliubov theory of
superconductivity~\cite{nnb71,nnb58,bts58}  was formulated on the
basis of a trial Hamiltonian  which  consists of a quadratic form
of creation and annihilation operators, including "anomalous" (off-diagonal) averages~\cite{bb09}.
The  strong-coupling BCS-Bogoliubov theory of superconductivity was  formulated for the Hubbard model
in the localized Wannier representation in Refs.~\cite{vkp82,khp83,wk83}.
Therefore, instead of the algebra of the normal state's
operator $a_{i\sigma}, a^{\dagger}_{i\sigma}$ and $n_{i\sigma}$, for description of superconducting
states, one has to use a more general algebra,
which includes the operators $a_{i\sigma}, a^{\dagger}_{i\sigma}, n_{i\sigma}$
and $a_{i\sigma}a_{i-\sigma}$, $a^{\dagger}_{i\sigma}a^{\dagger}_{i-\sigma}$.
The relevant generalized one-electron Green function will have  
the following form~\cite{rnc02,vkp82,khp83}:
\begin{eqnarray}
\label{e136}
 G_{ij} (\omega) =
\begin{pmatrix} 
G_{11}&G_{12}\cr 
G_{21}&G_{22}\cr 
\end{pmatrix}    =  \begin{pmatrix}
\langle \langle a_{i\sigma}\vert a^{\dagger}_{j\sigma} \rangle \rangle & \langle \langle a_{i\sigma}\vert
a_{j-\sigma} \rangle \rangle \cr \langle \langle a^{\dagger}_{i-\sigma}\vert
a^{\dagger}_{j\sigma} \rangle \rangle & \langle \langle a^{\dagger}_{i-\sigma}\vert
a_{j-\sigma} \rangle \rangle \cr    
\end{pmatrix}.
\end{eqnarray}
As it was  discussed in Refs.~\cite{kuz09,rnc02}, the off-diagonal (anomalous)
entries of the above matrix select the vacuum state of
the system in the BCS-Bogoliubov form, and they are
responsible for the presence of anomalous averages.
For treating the problem we follow  the general scheme of the irreducible Green functions 
method~\cite{kuz09,rnc02}.
In this approach we start from the equation of motion for the Green function $G_{ij} (\omega)$ (normal and anomalous components)
\begin{eqnarray}
\label{e137} \sum_{j}(\omega \delta_{ij} - t_{ij})\langle \langle a_{j\sigma}
\vert a^{\dagger}_{i'\sigma} \rangle \rangle = \delta_{ii'} + \\  \nonumber
U \langle \langle a_{i\sigma} n_{i-\sigma} \vert a^{\dagger}_{i'\sigma} \rangle \rangle +
\sum_{nj} V_{ijn} \langle \langle a_{j\sigma} u_{n} \vert a^{\dagger}_{i'\sigma} \rangle \rangle, \\
\label{e138} \sum_{j}(\omega \delta_{ij} +
t_{ij}) \langle \langle a^{\dagger}_{j-\sigma} \vert a^{\dagger}_{i'\sigma} \rangle \rangle =
\\  \nonumber
-U \langle \langle a^{\dagger}_{i-\sigma} n_{i\sigma} \vert
a^{\dagger}_{i'\sigma} \rangle \rangle + \sum_{nj} V_{jin}
\langle \langle a^{\dagger}_{j-\sigma} u_{n} \vert a^{\dagger}_{i'\sigma} \rangle \rangle.
\end{eqnarray}
The irreducible Green functions are introduced by definition
\begin{eqnarray}
\label{e139}
(^{(ir)}\langle \langle a_{i\sigma}a^{\dagger}_{i-\sigma}a_{i-\sigma} \vert
a^{\dagger}_{i'\sigma}\rangle \rangle_ {\omega} )   =
\langle \langle a_{i\sigma}a^{\dagger}_{i-\sigma}a_{i-\sigma}\vert
a^{\dagger}_{i'\sigma} \rangle \rangle_{\omega} - \\ \nonumber
- \langle n_{i-\sigma}\rangle G_{11} + \langle a_{i\sigma}a_{i-\sigma}\rangle
\langle \langle a^{\dagger}_{i-\sigma} \vert a^{\dagger}_{i'\sigma} \rangle \rangle_{\omega},\\
\nonumber
(^{(ir)}\langle \langle a^{\dagger}_{i\sigma}a_{i\sigma}a^{\dagger}_{i-\sigma}
\vert a^{\dagger}_{i'\sigma} \rangle \rangle_ {\omega} )  =
\langle \langle a^{\dagger}_{i\sigma}a_{i\sigma}a^{\dagger}_{i-\sigma}\vert
a^{\dagger}_{i'\sigma} \rangle \rangle_{\omega} - \\ \nonumber
- \langle n_{i\sigma}\rangle G_{21} +
\langle a^{\dagger}_{i\sigma}a^{\dagger}_{i-\sigma}\rangle \langle \langle a_{i\sigma} \vert
a^{\dagger}_{i'\sigma} \rangle \rangle_{\omega}. \nonumber
\end{eqnarray}
The self-consistent system of
superconductivity equations follows from the Dyson
equation~\cite{kuz09,rnc02,vkp82}
\begin{equation}
\label{e140} \hat G_{ii'}(\omega) = \hat G^{0}_{ii'}(\omega) +
\sum_{jj'} \hat G^{0}_{ij} (\omega) \hat M_{jj'}( \omega) \hat
G_{j'i'} (\omega).
\end{equation}
The mass operator $M_{jj'}( \omega)$ describes the processes of
inelastic electron scattering on lattice vibrations.
The elastic processes are described by the quantity (\ref{eq.64}).
Thus the "anomalous" off-diagonal terms fix the relevant
BCS-Bogoliubov vacuum and select the appropriate set of
solutions. The functional of the generalized mean
field   for the superconducting single-band Hubbard model  is of
the  form $\Sigma^{c}_{\sigma}$. 
A remark about the BCS-Bogoliubov mean-field
approach is instructive.  Speaking in physical terms, this theory
involves a condensation correctly, in spite that such a
condensation cannot be obtained by an expansion in the effective
interaction between electrons. Other mean field theories, {\it
e.g.} the Weiss molecular field theory and the van der Waals
theory of the liquid-gas transition are  much less reliable. The
reason why a mean-field theory of the superconductivity in the
BCS-Bogoliubov form is successful would appear to be that the
main correlations in metal are governed by the extreme degeneracy
of the electron gas. The correlations due to the pair
condensation, although they have dramatic effects, are weak (at
least in the ordinary superconductors) in comparison with the
typical electron energies, and may be treated in an average way
with a reasonable accuracy. 
It should be emphasized that
the high-temperature superconductors, discovered two decades ago, motivated an intensification  of
research in superconductivity, not only because applications are promising, but because they also represent a new state 
of matter that breaks certain fundamental symmetries. 
These are the broken symmetries of gauge (superconductivity), reflection ($d$-wave superconducting order parameter),
and time-reversal (ferromagnetism).\\
Superconductivity and antiferromagnetism are both the spontaneously broken symmetries. 
The question of symmetry breaking within the
localized and band models of antiferromagnets was
studied by the author of this work in Refs.~\cite{kuz09,km90,kuz99,kuz00}. It has been found there that the concept of
spontaneous symmetry breaking in the band model of
magnetism is much more complicated than in the
localized model. In the framework of the band model of
magnetism one has to additionally consider the so called
anomalous propagators of the form
\begin{eqnarray} \textrm{FM}: G_{fm} \sim \langle \langle a_{k\sigma};a^{\dag}_{k-\sigma} \rangle \rangle,  \nonumber \\
\nonumber \textrm{AFM}: G_{afm} \sim \langle \langle a_{k+Q\sigma};a^{\dag}_{k+Q'\sigma'} \rangle \rangle. \nonumber
\end{eqnarray}
In the case of the band antiferromagnet the ground
state of the system corresponds to a spin-density wave
(SDW), where a particle scattered on the internal inhomogeneous
periodic field gains the momentum $Q - Q'$
and changes its spin: $\sigma \rightarrow \sigma'$. The long-range order
parameters are defined as follows:
\begin{eqnarray} \label{e180}
\textrm{FM}: m =
1/N\sum_{k\sigma} \langle a^{\dag}_{k\sigma}a_{k-\sigma} \rangle,\\  \textrm{AFM}: M_{Q}
= \sum_{k\sigma} \langle a^{\dag}_{k\sigma}a_{k+Q-\sigma} \rangle. \label{e181}  \end{eqnarray}
It is important to stress, that the long-range order
parameters here are functionals of the internal field, which
in turn is a function of the order parameter. Thus, in the
cases of rotation and translation invariant Hamiltonians of
band ferro- and antiferromagnetics one has to add the following
infinitesimal sources removing the degeneracy:
\begin{eqnarray} \label{e182}
\textrm{FM}:  \nu\mu_{B}
H_{x}\sum_{k\sigma}a^{\dag}_{k\sigma}a_{k-\sigma},\\   \textrm{AFM}:
\nu \mu_{B} H \sum_{kQ} a^{\dag}_{k\sigma}a_{k+Q-\sigma}.
\label{e183}
\end{eqnarray}
Here, $\nu \rightarrow 0$ after the usual in statistical mechanics
infinite-volume limit $V \rightarrow \infty$. The ground state in the
form of a spin-density wave was obtained for the first
time by Overhauser. There, the vector  $\vec{Q}$ is a measure of inhomogeneity
or translation symmetry breaking in the system. 
The analysis performed  by various authors  showed  that the antiferromagnetic
and more complicated states (for instance, ferrimagnetic)
can be described in the framework of a generalized
mean-field approximation~\cite{kuz99}. In doing that we have to
take into account both the normal averages $\langle a^{\dag}_{i\sigma}a_{i\sigma}\rangle$
and the anomalous averages $\langle a^{\dag}_{i\sigma}a_{i-\sigma}\rangle$.  It is clear that
the anomalous terms  break the original
rotational symmetry of the Hubbard Hamiltonian.
Thus, the generalized mean-field's approximation for the antiferromagnet  has
the following form~\cite{kuz99} 
$n_{i-\sigma}a_{i\sigma} \simeq \langle n_{i-\sigma}\rangle a_{i\sigma} -
\langle a^{\dag}_{i-\sigma}a_{i\sigma}\rangle a_{i-\sigma}.$  
A self-consistent theory of band antiferromagnetism~\cite{kuz99}
was developed by the author of this work
using the method of the irreducible Green functions~\cite{kuz09,rnc02}. The
following definition of the irreducible Green functions  was used:
\begin{eqnarray} \label{e184}
^{ir}\langle \langle a_{k+p\sigma}a^{\dag}_{p+q-\sigma}a_{q-\sigma} \vert
a^{\dag}_{k\sigma} \rangle \rangle_ {\omega} =
\langle \langle a_{k+p\sigma}a^{\dag}_{p+q-\sigma}a_{q-\sigma}\vert
a^{\dag}_{k\sigma} \rangle \rangle_{\omega} - \nonumber\\ \delta_{p,
0}\langle n_{q-\sigma}\rangle G_{k\sigma} - \langle a_{k+p\sigma}a^{\dag}_{p+q-\sigma}\rangle
\langle \langle a_{q-\sigma} \vert a^{\dag}_{k\sigma} \rangle \rangle_{\omega}. \end{eqnarray} 
The algebra of relevant operators must be chosen as follows
($(a_{i\sigma}$,
$a^{\dag}_{i\sigma}$, $n_{i\sigma}$, $a^{\dag}_{i\sigma}a_{i-\sigma})$.
The corresponding
initial GF will have the following matrix structure
$$\mathcal{G}_{AFM} =  \begin{pmatrix}
\langle \langle a_{i\sigma}\vert a^{\dag}_{j\sigma}\rangle \rangle & \langle \langle a_{i\sigma}\vert
a^{\dag}_{j-\sigma}\rangle \rangle \cr
\langle \langle a_{i-\sigma}\vert a^{\dag}_{j\sigma}\rangle \rangle & \langle \langle a_{i-\sigma}\vert
a^{\dag}_{j-\sigma} \rangle \rangle \cr   \end{pmatrix}.$$
The off-diagonal terms select the vacuum state of
the band's antiferromagnet in the form of a spin-density
wave. 
With this definition, one
introduces the so-called anomalous (off-diagonal) Green functions which fix
the relevant vacuum and select the proper symmetry broken
solutions. 
The theory of the itinerant antiferromagnetism~\cite{kuz99} was formulated by using  sophisticated arguments of
the irreducible Green functions method in complete analogy with our description of the
Heisenberg  antiferromagnet at finite temperatures~\cite{km90}.
For the two-sublattice antiferromagnet we  used the matrix
Green function of the form
\begin{equation} \label{e92}
\hat G(k;\omega) = 
\begin{pmatrix}
   \langle \langle S^{+}_{ka} \vert S^{-}_{-ka} \rangle \rangle &
\langle \langle S^{+}_{ka} \vert S^{-}_{-kb}\rangle \rangle \cr \langle \langle S^{+}_{kb}\vert
S^{-}_{-ka}\rangle \rangle& \langle \langle S^{+}_{kb}\vert S^{-}_{-kb}\rangle \rangle \cr
\end{pmatrix}. 
\end{equation}
Here, the Green functions on the main diagonal are the usual or
normal Green functions, while the off-diagonal Green functions describe contributions
from the so-called anomalous terms, analogous
to the anomalous terms in the BCS-Bogoliubov superconductivity
theory. The anomalous (or off-diagonal)
average values in this case select the vacuum state
of the system precisely in the form of the two-sublattice
Neel state~\cite{kuz09}. 
%
%
\section{  Conclusions}
%
%
In the present work we shown that
the development and improvement of the methods of
quantum statistical mechanics still remains quite an
important direction of research~\cite{kuz09,kuz10}. 
In particular, the Bogoliubov's method of quasiaverages   gives the deep foundation and clarification of the
concept of broken symmetry~\cite{bsan}. 
It makes the emphasis on the notion of a degeneracy and
plays an important role in equilibrium statistical mechanics of many-particle systems. According to
that concept, infinitely small perturbations can trigger
macroscopic responses in the system if they break some
symmetry and remove the related degeneracy (or quasidegeneracy)
of the equilibrium state. As a result, they
can produce macroscopic effects even when the perturbation
magnitude is tend to zero, provided that happens
after passing to the thermodynamic limit. \\
We have discussed the theory of the
correlation effects for many-particle interacting systems  using
the ideas of   quasiaverages for   interacting electron
and spin systems on a lattice. The workable and self-consistent
\emph{irreducible Green functions} approach to the decoupling problem for the equation-of-motion
method for double-time temperature Green functions has been
presented.   The main advantage of the  formalism consists in the
clear separation of the elastic scattering corrections (generalized
mean fields) and inelastic scattering effects (damping and finite
lifetimes).  These effects could be self-consistently incorporated in a general
and compact manner.  Using   the IGF method, it is possible to obtain a closed
self-consistent set of equations determining the relevant Green functions and
self-energy. These equations
give a general microscopic description of correlation effects. Moreover,
this approach gives the workable scheme for the definition of
relevant generalized
mean fields written in terms of appropriate correlators.  
This  picture of interacting many-particle systems
on a lattice is far richer and gives more possibilities for the
analysis of phenomena which can actually take place. In this sense
the approach we described produces more advanced physical picture
of the quasiparticle many-body dynamics. 

\end{document}